\documentclass[10pt,a4paper,american,a4paper,aps,pra,onecolumn,notitlepage,longbibliography]{revtex4-1}
\usepackage{lmodern}
\usepackage[T1]{fontenc}
\usepackage[latin9]{inputenc}
\usepackage{array}
\usepackage{multirow}
\usepackage{amsmath}
\usepackage{amssymb}
\usepackage{graphicx}

\makeatletter

\providecommand{\tabularnewline}{\\}

\usepackage[colorlinks=true,linkcolor=blue,urlcolor=blue,citecolor=blue]{hyperref}

\makeatother

\usepackage{babel}
\begin{document}
\title{Spectral representation of Matsubara n-point functions:\\
Exact kernel functions and applications}
\author{Johannes Halbinger, Benedikt Schneider and Björn Sbierski}
\address{Department of Physics and Arnold Sommerfeld Center for Theoretical
Physics (ASC), Ludwig-Maximilians-Universität München, Theresienstr. 37,
München D-80333, Germany~\\
Munich Center for Quantum Science and Technology (MCQST), Schellingstr. 4,
D-80799 München, Germany}
\begin{abstract}
In the field of quantum many-body physics, the spectral (or Lehmann)
representation simplifies the calculation of Matsubara $n$-point
correlation functions if the eigensystem of a Hamiltonian is known.
It is expressed via a universal kernel function and a system- and
correlator-specific product of matrix elements. Here we provide the
kernel functions in full generality, for arbitrary $n$, arbitrary
combinations of bosonic or fermionic operators and an arbitrary number
of anomalous terms. As an application, we consider bosonic 3- and
4-point correlation functions for the fermionic Hubbard atom and a
free spin of length $S$, respectively.
\end{abstract}
\date{{\small{}\today}}
\maketitle

\section{Introduction and definitions\label{sec:Intro}}

Multi-point correlation functions of $n$ quantum mechanical operators,
also known as $n$-point functions, are a central concept in the study
of quantum many-body systems and field theory \citep{negeleQuantumManyParticle1988}.
They generalize the well-known 2-point functions, which, for the example
of electrons in the solid state, are routinely measured by scanning
tunneling spectroscopy or angle-resolved photon emission spectroscopy
\citep{bruusManyBodyQuantum2004}. For magnetic systems, the 2-point
spin correlators can be probed in a neutron scattering experiment.
Higher order correlation functions with $n=3,4,5...$ can for example
be measured in non-linear response settings \citep{kapplNonlinearResponses2022}.
In the emerging field of cold atomic quantum simulation, (equal-time)
$n$-point functions are even directly accessible \citep{semeghiniProbingTopological2021}. 

On the theoretical side the study of higher order correlation functions
gains traction as well. One motivation is the existence of exact relations
between correlation functions of different order $n$ \citep{hedinNewMethod1965,bickersSelfConsistentManyBody}.
Although these relations can usually not be solved exactly, they form
a valuable starting point for further methodological developments
like the parquet approximation \citep{rouletSingularitiesXRay1969}.
Thus even if the 4-point correlator (or, in that context, its essential
part, the one-line irreducible vertex \citep{negeleQuantumManyParticle1988})
might not be the primary quantity of interest in a calculation, it
appears as a building block of the method. Another example is the
functional renormalization group method (fRG) in a vertex expansion
\citep{kopietzIntroductionFunctional2010,metznerFunctionalRenormalization2012}.
It expresses the many body problem as a hierarchy of differential
equations for the vertices that interpolate between a simple solvable
starting point and the full physical theory \citep{wetterichExactEvolution1993}.
Whereas experiments measure correlation functions in real time (or
frequency), in theory one often is concerned with the related but
conceptually simpler versions depending on imaginary time \citep{negeleQuantumManyParticle1988}.
In the following, we will focus on these Matsubara correlation functions,
which, nevertheless feature an intricate frequency dependence.

Whereas the above theoretical methods usually provide only an approximation
for the $n$-point functions, an important task is to calculate these
objects exactly. This should be possible for simple quantum many body
systems. We consider systems simple if they are amenable to exact
diagonalization (ED), i.e.~feature a small enough Hilbert space,
like few-site clusters of interacting quantum spins or fermions. Also
impurity systems, where interactions only act locally, can be approximately
diagonalized using the numerical renormalization group \citep{leeComputingLocal2021}.

Knowing the exact $n$-point functions for simple systems is important
for benchmark testing newly developed methods before deploying them
to harder problems. Moreover, $n$-point functions for simple systems
often serve as the starting point of further approximations like in
the spin-fRG \citep{reutherClusterFunctional2014,kriegExactRenormalization2019,ruckriegelSpinFunctional2022},
or appear intrinsically in a method like in diagrammatic extensions
of dynamical mean field theory \citep{rohringerDiagrammaticRoutes2018}
with its auxiliary impurity problems. Another pursuit enabled by the
availability of exact $n$-point functions is to interpret the wealth
of information encoded in these objects, in particular in their rich
frequency structure. For example, Ref.~\citep{chalupaFingerprintsLocal2021}
studied the fingerprints of local moment formation and Kondo screening
in quantum impurity models.

\begin{figure}[b]
\begin{centering}
\includegraphics{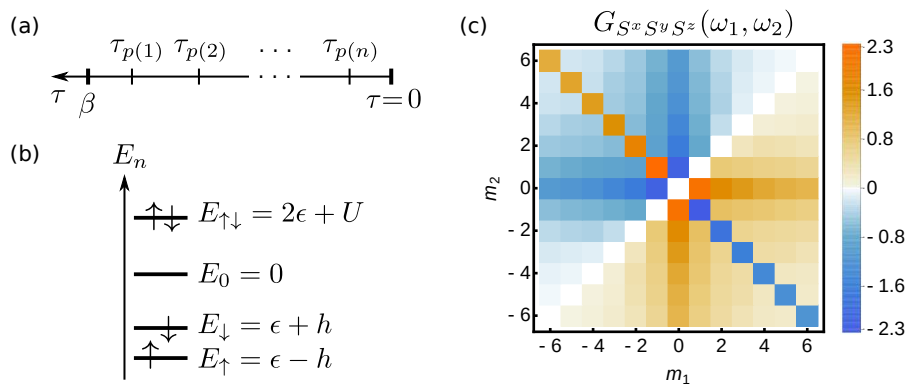}
\par\end{centering}
\caption{\label{fig:fig}(a) Ordering convention for imaginary times in Eq.~(\ref{eq:GwithP}).
(b) Eigenstates and energies of the Hubbard atom. (c) Matsubara correlation
function $G_{S^{x}S^{y}S^{z}}(\omega_{1},\omega_{2})$ with $\omega_{j}=2\pi m_{j}/\beta$
($m_{j}\in\mathbb{Z}$, $j=1,2$) for the Hubbard atom (\ref{eq:HAHamiltonian})
at $\beta=10$, $h=0.1$, $\epsilon=-2$, $U=2$, see Eq.~(\ref{eq:GSxSySz}).
The sharp anti-diagonal ray $\propto\delta_{\omega_{1}+\omega_{2},0}$
represents an anomalous term of order $a=1$. The other broadened
rays become sharp and anomalous for $h\rightarrow0$, see Eq.~(\ref{eq:GSxSySz(h=00003D0)}).}
\end{figure}

In this work we complete the task to calculate exact $n$-point functions
by generalizing the spectral (or Lehmann) representation \citep{lehmannUeberEigenschaften1954,negeleQuantumManyParticle1988}
for Matsubara $n$-point correlation functions to arbitrary $n$.
We assume that a set of eigenstates and -energies is given. Following
pioneering work of Refs.~\citep{shvaika3pt2006,hafermannSuperperturbationSolver2009,shvaika4pt2016}
and in particular the recent approach by Kugler \emph{et al.}~\citep{kuglerMultipointCorrelation2021},
we split the problem of calculating imaginary frequency correlators
into the computation of a universal kernel function and a system-
and correlator-specific part (called partial spectral function in
Ref.~\citep{kuglerMultipointCorrelation2021}). We provide the kernel
functions in full generality for an arbitrary number $n$ of bosonic
or fermionic frequencies. Previously, these kernel functions were
known exactly only up to the 3-point case \citep{shvaika3pt2006},
for the fermionic 4-point case \citep{hafermannSuperperturbationSolver2009,shvaika4pt2016,kuglerMultipointCorrelation2021}
or for the general $n$-point case \citep{kuglerMultipointCorrelation2021}
but disregarding anomalous contributions to the sum that the kernel
function consists of. These anomalous contributions are at the heart
of the complexity of Matsubara $n$-point functions. They occur when
certain combinations of eigenenergies and external frequencies vanish
individually, see the anti-diagonal rays in Fig.~\ref{fig:fig}(c).
Physically, they correspond to long-term memory effects, are related
to non-ergodicity and, in the case of bosonic two-point functions
reflect the difference between static isothermal susceptibilities
and the zero-frequency limit of the dynamical Kubo response function
\citep{kwokCorrelationFunctions1969,watzenbockLongtermMemory2022}.

The structure of the paper is as follows: In Sec.~\ref{sec:Def}
we define the Matsubara $n$-point function $G_{A_{1}...A_{n}}\bigl(\omega_{1},...,\omega_{n-1}\bigr)$
and review some of its properties. The spectral representation is
derived in Sec.~\ref{sec:Spectral-rep} with Eq.~(\ref{eq:Spectral_Gw})
being the central equation written in terms of the kernel function
$K_{n}(\Omega_{1},...,\Omega_{n-1})$. Our main result is an exact
closed-form expression of this most general kernel function which
is given in Sec.~\ref{sec:exact-expression}. Examples for $n=2,3,4,5$
are given in Sec.~\ref{sec:n2345Kernels} where we also discuss simplifications
for the purely fermionic case. We continue with applications to two
particular systems relevant in the field of condensed matter theory:
In Sec.~\ref{sec:Applications}, we consider the Hubbard atom and
the free spin of length $S$, for which we compute $n$-point functions
not previously available in the literature. We conclude in Sec.~\ref{sec:Conclusion}.

\section{Definition of Matsubara $n$-point function $G_{A_{1}...A_{n}}\bigl(\omega_{1},...,\omega_{n}\bigr)$\label{sec:Def}}

We consider a set of $n=2,3,4,...$ operators $\{A_{1},A_{2},...,A_{n}\}$
defined on the Hilbert space of a quantum many-body Hamiltonian $H$.
The operators can be fermionic, bosonic or a combination of both types,
with the restriction that there is an even number of fermionic operators.
As an example, $A_{1}=d^{\dagger}d\equiv n$, $A_{2}=d$, $A_{3}=d^{\dagger}$
where $d^{\dagger}$ and $d$ are canonical fermionic creation and
annihilation operators. A subset of operators is called bosonic if
they create a closed algebra under the commutation operation. They
are called fermionic if the algebra is closed under anti-commutation,
see Sec.~1 of Ref.~\citep{tsvelikQuantumField2007}. Spin operators
are thus bosonic.

We define the imaginary time-ordered $n$-point correlation functions
for $\tau_{k}\in[0,\beta]$ \citep{rohringerLocalElectronic2012,rohringerNewRoutes2013},
\begin{equation}
G_{A_{1}A_{2}...A_{n}}\left(\tau_{1},\tau_{2},...,\tau_{n}\right)\equiv\left\langle \mathcal{T}A_{1}(\tau_{1})A_{2}(\tau_{2})...A_{n}(\tau_{n})\right\rangle ,\label{eq:G(tau)}
\end{equation}
where $A_{k}(\tau_{k})=e^{\tau_{k}H}A_{k}e^{-\tau_{k}H}$ denotes
Heisenberg time evolution. Here and in the following, $k=1,2,...,n$.
The expectation value is calculated as $\bigl\langle...\bigr\rangle=\mathrm{tr}[\rho...]$
where $\rho=\exp(-\beta H)/Z$ is the thermal density operator at
temperature $\beta=1/T$ and $Z=\mathrm{tr}\exp(-\beta H)$ denotes
the partition function. Note that other conventions for the $n$-point
function differing by a prefactor are also used in the literature,
e.g.~Ref.~\citep{kuglerMultipointCorrelation2021} multiplies with
$(-1)^{n-1}$. In Eq.~(\ref{eq:G(tau)}), the imaginary time-ordering
operator $\mathcal{T}$ orders the string of Heisenberg operators,
\begin{equation}
\mathcal{T}A_{1}(\tau_{1})A_{2}(\tau_{2})...A_{n}(\tau_{n})\equiv\boldsymbol{\zeta}(p)A_{p(1)}(\tau_{p(1)})A_{p(2)}(\tau_{p(2)})...A_{p(n)}(\tau_{p(n)}),
\end{equation}
where $p$ is the permutation $p\in S_{n}$ such that $\tau_{p(1)}>\tau_{p(2)}>...>\tau_{p(n)}$
{[}see Fig.~\ref{fig:fig}(a){]} and the sign $\boldsymbol{\zeta}(p)$
is $-1$ if the operator string $A_{p(1)}A_{p(2)}...A_{p(n)}$ differs
from $A_{1}A_{2}...A_{n}$ by an odd number of transpositions of fermionic
operators, otherwise it is $+1$. The special case $n=2$, with $\boldsymbol{\zeta}(12)=1$
and $\boldsymbol{\zeta}(21)=\zeta$ ($\zeta=1$ for $A_{1,2}$ bosonic,
$\zeta=-1$ for $A_{1,2}$ fermionic), simplifies to 
\begin{equation}
\mathcal{T}A_{1}(\tau_{1})A_{2}(\tau_{2})=\begin{cases}
A_{1}(\tau_{1})A_{2}(\tau_{2}) & :\tau_{1}>\tau_{2},\\
\zeta A_{2}(\tau_{2})A_{1}(\tau_{1}) & :\tau_{2}>\tau_{1}.
\end{cases}
\end{equation}

Imaginary time-ordered correlation functions (\ref{eq:G(tau)}) fulfill
certain properties which we review in the following, see e.g.~\citep{rohringerNewRoutes2013}
for a more extensive discussion. First, they are invariant under translation
of all time arguments,
\begin{equation}
G_{A_{1}A_{2}...A_{n}}\left(\tau_{1},\tau_{2},...,\tau_{n}\right)=G_{A_{1}A_{2}...A_{n}}\left(\tau_{1}+\tau,\tau_{2}+\tau,...,\tau_{n}+\tau\right),\label{eq:tau-transl}
\end{equation}
with $\tau\in\mathbb{R}$ such that $\tau_{k}+\tau\in[0,\beta]$.
They also fulfill periodic or anti-periodic boundary conditions for
the individual arguments $\tau_{k}$,
\begin{equation}
G_{A_{1}...A_{n}}\left(\tau_{1},...,\tau_{k}=0,...,\tau_{n}\right)=\zeta_{k}G_{A_{1}...A_{n}}\left(\tau_{1},...,\tau_{k}=\beta,...,\tau_{n}\right)\label{eq:G(tau)BC}
\end{equation}
where $\zeta_{k}=+1$ or $-1$ if $A_{k}$ is from the bosonic or
fermionic subset of operators, respectively. This motivates the use
of a Fourier transformation,

\begin{eqnarray}
G_{A_{1}...A_{n}}\left(\tau_{1},...,\tau_{n}\right) & \equiv & \beta^{-n}\sum_{\omega_{1},...,\omega_{n}}e^{-i(\omega_{1}\tau_{1}+...+\omega_{n}\tau_{n})}G_{A_{1}...A_{n}}\left(\omega_{1},...,\omega_{n}\right),\\
G_{A_{1}...A_{n}}\left(\omega_{1},...,\omega_{n}\right) & = & \int_{0}^{\beta}\mathrm{d}\tau_{1}\cdots\int_{0}^{\beta}\mathrm{d}\tau_{n}e^{+i(\omega_{1}\tau_{1}+...+\omega_{n}\tau_{n})}G_{A_{1}...A_{n}}\left(\tau_{1},...,\tau_{n}\right),\label{eq:G(w)}
\end{eqnarray}
where $\omega_{k}=2\pi m_{k}/\beta$ or $\omega_{k}=2\pi(m_{k}+1/2)/\beta$
with $m_{k}\in\mathbb{Z}$ bosonic or fermionic Matsubara frequencies,
respectively, and $\sum_{\omega_{k}}$ is shorthand for $\sum_{m_{k}\in\mathbb{Z}}$.
Note that fermionic Matsubara frequencies are necessarily nonzero,
a property that will become important later. As we will not discuss
the real-frequency formalisms, we will not write the imaginary unit
in front of Matsubara frequencies in the arguments of $G_{A_{1}...A_{n}}(\omega_{1},...,\omega_{n})$.
Again, note that in the literature, different conventions for the
Fourier transformation of $n$-point functions are in use. In particular
some authors pick different signs in the exponent of Eq.~(\ref{eq:G(w)})
for fermionic creation and annihilation operators, or chose these
signs depending on operator positions.

Time translational invariance (\ref{eq:tau-transl}) implies frequency
conservation at the left hand side of Eq.~(\ref{eq:G(w)}),
\begin{equation}
G_{A_{1}...A_{n}}\bigl(\omega_{1},...,\omega_{n-1},\omega_{n}\bigr)\equiv\beta\delta_{0,\omega_{1}+...+\omega_{n}}G_{A_{1}...A_{n}}\bigl(\omega_{1},...,\omega_{n-1}\bigr),\label{eq:beta_delta_G}
\end{equation}
where on the right hand side we skipped the $n$-th frequency entry
in the argument list of $G$. Note that we do not use a new symbol
for the correlation function when we pull out the factor $\beta$
and the Kronecker delta function. 

\section{Spectral representation of $G_{A_{1}...A_{n}}\bigl(\omega_{1},...,\omega_{n-1}\bigr)$\label{sec:Spectral-rep}}

The integrals involved in the Fourier transformation (\ref{eq:G(w)})
generate all $n!$ different orderings of the time arguments $\tau_{k}$.
As in Ref.~\citep{kuglerMultipointCorrelation2021} it is thus convenient
to use a sum over all $n!$ permutations $p\in S_{n}$ and employ
a product of $n-1$ step-functions $\theta$, with $\theta(x)=1$
for $x>0$ and $0$ otherwise, to filter out the unique ordering for
which $\beta>\tau_{p(1)}>\tau_{p(2)}>...>\tau_{p(n-1)}>\tau_{p(n)}>0$,
see Fig.~\ref{fig:fig}(a),
\begin{eqnarray}
G_{A_{1}...A_{n}}(\tau_{1},...,\tau_{n}) & = & \sum_{p\in S_{n}}\boldsymbol{\zeta}(p)\left[\prod_{i=1}^{n-1}\theta(\tau_{p(i)}-\tau_{p(i+1)})\right]\left\langle A_{p(1)}(\tau_{p(1)})A_{p(2)}(\tau_{p(2)})...A_{p(n)}(\tau_{p(n)})\right\rangle .\label{eq:GwithP}
\end{eqnarray}
To expose explicitly the time dependence of the Heisenberg operators,
we insert $n$ times the basis of eigenstates and -energies of the
many-body Hamiltonian $H$. Instead of the familiar notation $\bigl|j_{1}\bigr\rangle,\bigl|j_{2}\bigr\rangle,...$
and $E_{j_{1}},E_{j_{2}},...$ we employ $\bigl|\underline{1}\bigr\rangle,\bigl|\underline{2}\bigr\rangle,...$
and $E_{\underline{1}},E_{\underline{2}}$,.... for compressed notation
and denote operator matrix elements as $A^{\underline{1}\underline{2}}=\left\langle \underline{1}|A|\underline{2}\right\rangle $.
We obtain
\begin{eqnarray}
 &  & G_{A_{1}...A_{n}}(\tau_{1},...,\tau_{n})=\sum_{p\in S_{n}}\boldsymbol{\zeta}(p)\left[\prod_{i=1}^{n-1}\theta(\tau_{p(i)}-\tau_{p(i+1)})\right]\\
 &  & \times\frac{1}{Z}\sum_{\underline{1}...\underline{n}}e^{-\beta E_{\underline{1}}}e^{\tau_{p(1)}E_{\underline{1}}}A_{p(1)}^{\underline{1}\,\underline{2}}e^{(-\tau_{p(1)}+\tau_{p(2)})E_{\underline{2}}}A_{p(2)}^{\underline{2}\,\underline{3}}e^{(-\tau_{p(2)}+\tau_{p(3)})E_{\underline{3}}}...e^{(-\tau_{p(n-1)}+\tau_{p(n)})E_{\underline{n}}}A_{p(n)}^{\underline{n}\,\underline{1}}e^{-\tau_{p(n)}E_{\underline{1}}},\nonumber 
\end{eqnarray}
and apply the Fourier transform according to the definition~(\ref{eq:G(w)}),
\begin{eqnarray}
 &  & G_{A_{1}...A_{n}}\bigl(\omega_{1},...,\omega_{n}\bigr)=\frac{1}{Z}\sum_{p\in S_{n}}\boldsymbol{\zeta}(p)\sum_{\underline{1}...\underline{n}}e^{-\beta E_{\underline{1}}}A_{p(1)}^{\underline{1}\,\underline{2}}A_{p(2)}^{\underline{2}\,\underline{3}}...A_{p(n)}^{\underline{n}\,\underline{1}}\label{eq:preKernel}\\
 &  & \times\left[\int_{0}^{\beta}\!\!\mathrm{d}\tau_{p(1)}e^{\Omega_{p(1)}^{\underline{1}\,\underline{2}}\tau_{p(1)}}\right]\left[\int_{0}^{\tau_{p(1)}}\!\!\mathrm{d}\tau_{p(2)}e^{\Omega_{p(2)}^{\underline{2}\,\underline{3}}\tau_{p(2)}}\right]...\left[\int_{0}^{\tau_{p(n-2)}}\!\!\mathrm{d}\tau_{p(n-1)}e^{\Omega_{p(n-1)}^{\underline{n-1}\,\underline{n}}\tau_{p(n-1)}}\right]\left[\int_{0}^{\tau_{p(n-1)}}\!\!\mathrm{d}\tau_{p(n)}e^{\Omega_{p(n)}^{\underline{n}\,\underline{1}}\tau_{p(n)}}\right],\nonumber 
\end{eqnarray}
where we defined 

\begin{equation}
\Omega_{k}^{\underline{a}\,\underline{b}}\equiv i\omega_{k}+E_{\underline{a}}-E_{\underline{b}}\in\mathbb{C}.\label{eq:Omega_k^ab}
\end{equation}
In Eq.~(\ref{eq:preKernel}), the first line carries all the information
of the system and the set of operators $\{A_{1},A_{2},...,A_{n}\}$.
The second line can be regarded as a universal kernel function defined
for general $\{\Omega_{1},\Omega_{2},...,\Omega_{n}\}$ probed at
$\Omega_{k}\in\mathbb{C}$ which depends on the system and correlators
via (\ref{eq:Omega_k^ab}). Upon renaming the $\tau$-integration
variables $\tau_{p(k)}\rightarrow\tau_{k}$, this kernel function
is written as follows:

\begin{eqnarray}
\mathcal{K}_{n}\left(\Omega_{1},...,\Omega_{n}\right) & \equiv & \left[\int_{0}^{\beta}\mathrm{d}\tau_{1}e^{\Omega_{1}\tau_{1}}\right]\left[\int_{0}^{\tau_{1}}\mathrm{d}\tau_{2}e^{\Omega_{2}\tau_{2}}\right]...\left[\int_{0}^{\tau_{n-2}}\mathrm{d}\tau_{n-1}e^{\Omega_{n-1}\tau_{n-1}}\right]\left[\int_{0}^{\tau_{n-1}}\mathrm{d}\tau_{n}e^{\Omega_{n}\tau_{n}}\right]\label{eq:calKIncr}\\
 & \equiv & \beta\delta_{0,\Omega_{1}+\Omega_{2}+...+\Omega_{n}}K_{n}\left(\Omega_{1},...,\Omega_{n-1}\right)+R_{n}\left(\Omega_{1},...,\Omega_{n}\right).\label{eq:KIncr}
\end{eqnarray}
In the second line we split $\mathcal{K}_{n}$ into a part $K_{n}$
proportional to $\beta\delta_{0,\Omega_{1}+\Omega_{2}+...+\Omega_{n}}$
and the rest $R_{n}$. We dropped $\Omega_{n}$ from the argument
list of $K_{n}$ which can be reconstructed from $\{\Omega_{1},...,\Omega_{n-1}\}$.

Finally, we express $G_{A_{1}...A_{n}}\bigl(\omega_{1},...,\omega_{n}\bigr)$
of Eq.~(\ref{eq:preKernel}) using the kernel $\mathcal{K}_{n}$
so that the general $\Omega_{k}\in\mathbb{C}$ get replaced by $\Omega_{k}^{\underline{a}\,\underline{b}}$
of Eq.~(\ref{eq:Omega_k^ab}). For these, $\Omega_{p(1)}^{\underline{1}\,\underline{2}}+\Omega_{p(2)}^{\underline{2}\,\underline{3}}+...+\Omega_{p(n)}^{\underline{n}\,\underline{1}}=i(\omega_{1}+\omega_{2}+...+\omega_{n})$,
since the $E_{\underline{k}}$ cancel pairwise. The structure of Eq.~(\ref{eq:beta_delta_G})
(which followed from time translational invariance) implies that the
terms proportional to $R_{n}$ are guaranteed to cancel when summed
over permutations $p\in S_{n}$, so that only the terms proportional
to $K_{n}$ remain. We drop the $\beta\delta_{0,\omega_{1}+\omega_{2}+...+\omega_{n}}$
from both sides {[}c.f.~Eq.~(\ref{eq:beta_delta_G}){]} and find
the spectral representation of the $n$-point correlation function
in the Matsubara formalism,
\begin{equation}
\boxed{G_{A_{1}...A_{n}}\bigl(\omega_{1},...,\omega_{n-1}\bigr)=\frac{1}{Z}\sum_{p\in S_{n}}\boldsymbol{\zeta}(p)\sum_{\underline{1}...\underline{n}}e^{-\beta E_{\underline{1}}}A_{p(1)}^{\underline{1}\,\underline{2}}A_{p(2)}^{\underline{2}\,\underline{3}}...A_{p(n)}^{\underline{n}\,\underline{1}}\times K_{n}\left(\Omega_{p(1)}^{\underline{1}\,\underline{2}},\Omega_{p(2)}^{\underline{2}\,\underline{3}},...,\Omega_{p(n-1)}^{\underline{n-1}\,\underline{n}}\right).}\label{eq:Spectral_Gw}
\end{equation}
An equivalent expression was derived in the literature before \citep{kuglerMultipointCorrelation2021},
see also Refs.~\citep{shvaika3pt2006,hafermannSuperperturbationSolver2009,shvaika4pt2016}
for the cases of certain small $n$. However, the kernel functions
$K_{n}$ where previously only known approximately, for situations
involving only a low order of anomalous terms, see the discussion
in Sec.~\ref{sec:n2345Kernels}. We define an \emph{anomalous} term
of order $a=1,2,...n-1$ as a summand contributing to $K_{n}\left(\Omega_{1},...,\Omega_{n-1}\right)$
that contains a product of $a$ Kronecker delta functions $\delta_{0,x}$,
where $x$ is a sum of a subset of $\{\Omega_{1},...,\Omega_{n-1}\}$.
As can be seen in Fig.~\ref{fig:fig}(c), these anomalous contributions
to $G_{A_{1}...A_{n}}\bigl(\omega_{1},...,\omega_{n}\bigr)$ correspond
to qualitatively important sharp features. 

In the next section, we present a simple, exact expression for general
$K_{n}\left(\Omega_{1},...,\Omega_{n-1}\right)$. Readers not interested
in the derivation can directly skip to the result in Eq.~(\ref{eq:KnFinal})
or its explicit form for $n=2,3,4,5$ in Sec.~(\ref{sec:n2345Kernels}).

\section{General kernel function $K_{n}\bigl(\Omega_{1},...,\Omega_{n-1}\bigr)$\label{sec:exact-expression}}

Assuming the spectrum and matrix elements entering Eq.~(\ref{eq:Spectral_Gw})
are known, the remaining task is to find expressions for the kernel
function $K_{n}\left(\Omega_{1},...,\Omega_{n-1}\right)$ defined
via Eqns.~(\ref{eq:calKIncr}) and (\ref{eq:KIncr}) as the part
of $\mathcal{K}_{n}\left(\Omega_{1},\Omega_{2},...,\Omega_{n}\right)$
multiplying $\beta\delta_{0,\Omega_{1}+\Omega_{2}+...+\Omega_{n}}$.
To facilitate the presentation in this section, in Eq.~(\ref{eq:calKIncr})
we rename the integration variables $\tau_{k}\rightarrow\tau_{n-k+1}$
and define new arguments $z_{n-j+1}=\Omega_{j}$ for $j=1,2,...,n-1$,
\begin{align}
\mathcal{K}_{n}\left(\Omega_{1}=z_{n},\Omega_{2}=z_{n-1},...,\Omega_{n}=z_{1}\right) & =\Bigl[\int_{0}^{\beta}\mathrm{d}\tau_{n}e^{z_{n}\tau_{n}}\Bigr]\Bigl[\int_{0}^{\tau_{n}}\mathrm{d}\tau_{n-1}e^{z_{n-1}\tau_{n-1}}\Bigr]...\Bigl[\int_{0}^{\tau_{3}}\mathrm{d}\tau_{2}\underset{\equiv h_{2}(\tau_{2})}{\underbrace{e^{z_{2}\tau_{2}}\Bigr]\Bigl[\int_{0}^{\tau_{2}}\mathrm{d}\tau_{1}\underset{\equiv h_{1}(\tau_{1})}{\underbrace{e^{z_{1}\tau_{1}}}}\Bigr]}}\label{eq:calKDecr}\\
 & =\beta\delta_{0,z_{1}+z_{2}+...+z_{n}}K_{n}\left(z_{n},z_{n-1},...,z_{2}\right)+R\left(z_{n},z_{n-1},...,z_{1}\right).\label{eq:KDecr}
\end{align}
As indicated in Eq.~(\ref{eq:calKDecr}), we call $h_{k}(\tau_{k})$
the integrand of the $\int_{0}^{\tau_{k+1}}\mathrm{d}\tau_{k}$ integral
for $k=1,2,...,n$. At $k=1$ this integrand is given by $h_{1}(\tau_{1})=e^{z_{1}\tau_{1}}$
and we will find $h_{k}$ for $k=2,3,...,n$ iteratively. For $z\in\mathbb{C}$,
we define the abbreviations $\delta_{z}\equiv\delta_{0,z}$ and
\begin{equation}
\Delta_{z}\equiv\begin{cases}
0 & \textnormal{{if} }z=0\\
\frac{{1}}{z} & \textnormal{{if} }z\neq0
\end{cases}\label{eq:Deltaz}
\end{equation}
and consider the integral (for $p=0,1,2,...$ and $\tilde{\tau}\geq0$,
proof by partial integration and induction)
\begin{equation}
\int_{0}^{\tilde{\tau}}\mathrm{d}\tau\,\tau^{p}e^{z\tau}=\left[\frac{\tilde{\tau}^{p+1}}{p+1}\delta_{z}+p!\left(-1\right)^{p}\Delta_{z}^{1+p}\sum_{l=0}^{p}\frac{(-1)^{l}}{l!}\Delta_{z}^{-l}\tilde{\tau}^{l}\right]e^{z\tilde{\tau}}-p!\left(-1\right)^{p}\Delta_{z}^{p+1}.\label{eq:basicInt}
\end{equation}
Recall that we are only interested in the contribution $K_{n}\left(z_{n},z_{n-1},...,z_{2}\right)$
that fulfills frequency conservation, see Eq.~(\ref{eq:KDecr}).
The $\delta_{z_{1}+z_{2}+...+z_{n}}$ in front of this term arises
from the final $\tau_{n}$ integration of $h_{n}(\tau_{n})\propto e^{(z_{1}+z_{2}+...+z_{n})\tau_{n}}$
via the first term in Eq.~(\ref{eq:basicInt}). This however requires
that all $z_{k}$ (except the vanishing ones, of course) remain in
the exponent during the iterative integrations. This requirement is
violated by the last term in the general integral (\ref{eq:basicInt})
(which comes from the lower boundary of the integral). All terms in
$\mathcal{K}_{n}$ that stem from this last term in Eq.~(\ref{eq:basicInt})
thus contribute to $R_{n}$ and can be dropped in the following \citep{kuglerMultipointCorrelation2021}.
Note however, that it is straightforward to generalize our approach
and keep these terms if the full $\mathcal{K}_{n}$ is required.

To define the iterative procedure to solve the $n$-fold integral
in Eq.~(\ref{eq:calKDecr}), we make the ansatz 
\begin{equation}
h_{k}(\tau_{k})=\sum_{l=0}^{k-1}f_{k}(l)\tau_{k}^{l}e^{\left(z_{k}+z_{k-1}+...+z_{1}\right)\tau_{k}},\label{eq:h_k-Ansatz}
\end{equation}
which follows from the form of the integral (\ref{eq:basicInt}) and
our decision to disregard the terms contributing to $R_{n}$. The
ansatz $\eqref{eq:h_k-Ansatz}$ is parameterized by the numbers $f_{k}(l)$
with $l=0,1,...,k-1$. These numbers have to be determined iteratively,
starting from $f_{k=1}(l=0)=1$, read off from $h_{1}(\tau_{1})=e^{z_{1}\tau_{1}}$,
c.f.~Eq.~(\ref{eq:calKDecr}). Iteration rules to obtain the $f_{k}(l)$
from $f_{k-1}(l)$ are easily derived from Eqns.~(\ref{eq:calKDecr}),
(\ref{eq:basicInt}) and $\eqref{eq:h_k-Ansatz}$. We obtain the recursion
relation
\begin{equation}
f_{k}(l)=\sum_{p=0}^{k-1}\tilde{M}_{k-1}(l,p)f_{k-1}(p)\label{eq:fIt}
\end{equation}
written as a matrix-vector product of $\mathbf{f}_{k-1}=(f_{k-1}(0),f_{k-1}(1),...,f_{k-1}(k-2))^{\mathrm{T}}$
with the $k\times(k-1)$-matrix 
\begin{equation}
\tilde{M}_{k-1}(l,p)=\frac{p!}{l!}\left[\delta_{l,p+1}\tilde{\delta}_{k-1}+\theta\left(p-l+1/2\right)\left(-1\right)^{l+p}\tilde{\Delta}_{k-1}^{1+p-l}\right],\label{eq:Mtilde}
\end{equation}
where $\tilde{\Delta}_{k}\equiv\Delta_{z_{k}+...+z_{2}+z_{1}}$, $\tilde{\delta}_{k}\equiv\delta_{z_{k}+...+z_{2}+z_{1}}$.
The tilde on top of the $\tilde{\delta}_{k}$ and $\tilde{\Delta}_{k}$
signals the presence of a sum of $z_{j}$ in the arguments (below
we will define related quantities without tilde for the sum of $\Omega_{j}$).
Note that the first (second) term in brackets of Eq.~(\ref{eq:Mtilde})
comes from the first (second) term in square brackets of Eq.~(\ref{eq:basicInt}).

The next step is to find $K_{n}\left(z_{n},z_{n-1},...,z_{2}\right)$.
This requires to do the integral $\int_{0}^{\beta}\mathrm{d}\tau_{n}h_{n}(\tau_{n})$
which can be again expressed via Eq.~(\ref{eq:basicInt}) but with
the replacement $\tilde{\tau}\rightarrow\beta$. Only the first term
provides a $\beta\delta_{z_{1}+z_{2}+...+z_{n}}$ and is thus identified
with $K_{n}$. We find:
\begin{equation}
K_{n}\left(z_{n},z_{n-1},...,z_{2}\right)=\sum_{l=0}^{n-1}\frac{\beta^{l}f_{n}(l)}{l+1}.\label{eq:KnFromfn}
\end{equation}
The argument $z_{1}$ that the right hand side of Eq.~(\ref{eq:KnFromfn})
depends on is to be replaced by $z_{1}=-z_{2}-z_{3}-...-z_{n}$, in
line with the arguments in $K_{n}\left(z_{n},z_{n-1},...,z_{2}\right)$.
Then, to conform with Eq.~(\ref{eq:Spectral_Gw}), we reinstate $\Omega_{j}=z_{n-j+1}$
for $j=1,2,...,n-1$. This amounts to replacing the terms $\tilde{\delta}_{j}$
and $\tilde{\Delta}_{j}$ that appear in $f_{n}(l)$ as follows,
\begin{eqnarray}
\tilde{\delta}_{j} & = & \delta_{z_{j}+...+z_{2}+z_{1}}=\delta_{\Omega_{1}+\Omega_{2}+...+\Omega_{n-j}}\equiv\delta_{n-j},\label{eq:delta_n-j}\\
-\tilde{\Delta}_{j} & = & -\Delta_{z_{j}+...+z_{2}+z_{1}}=\Delta_{\Omega_{1}+\Omega_{2}+...+\Omega_{n-j}}\equiv\Delta_{n-j},\label{eq:Delta_n-j}
\end{eqnarray}
where we used $\Omega_{1}+\Omega_{2}+...+\Omega_{n}=0=z_{n}+...+z_{2}+z_{1}$.
Finally, we can express Eq.~(\ref{eq:KnFromfn}) using a product
of $n-1$ matrices $\tilde{M}$ multiplying the initial length-1 vector
with entry $f_{1}(0)=1$. Transferring to the $\Omega$-notation by
using Eqns.~(\ref{eq:delta_n-j}) and (\ref{eq:Delta_n-j}), we obtain
\begin{equation}
\boxed{K_{n}\left(\Omega_{1},...,\Omega_{n-1}\right)=\!\!\!\sum_{i_{n-1}=0}^{n-1}\sum_{i_{n-2}=0}^{n-2}\!\cdots\!\sum_{i_{2}=0}^{2}\sum_{i_{1}=0}^{1}\frac{\beta^{i_{n-1}}}{i_{n-1}+1}M_{1}(i_{n-1},i_{n-2})M_{2}(i_{n-2},i_{n-3})\cdots M_{n-2}(i_{2},i_{1})M_{n-1}(i_{1},0)}\label{eq:KnFinal}
\end{equation}
with
\begin{equation}
M_{j}(l,p)\equiv\frac{p!}{l!}\left[\delta_{l,p+1}\delta_{j}-\theta\left(p-l+1/2\right)\Delta_{j}^{1+p-l}\right].\label{eq:M}
\end{equation}
The closed form expression (\ref{eq:KnFinal}) of the universal kernel,
to be used in the spectral representation (\ref{eq:Spectral_Gw}),
is our main result. By definition it is free of any singularities
as the case of vanishing denominators is explicitly excluded in Eq.~(\ref{eq:Deltaz}).

\section{Explicit kernel functions $K_{n}\bigl(\Omega_{1},...,\Omega_{n-1}\bigr)$
for $n=2,3,4,5$\label{sec:n2345Kernels}}

While the previous section gives a closed form expression for kernel
functions of arbitrary order, we here evaluate the universal kernel
functions $K_{n}\left(\Omega_{1},...,\Omega_{n-1}\right)$ defined
in Eq.~(\ref{eq:KIncr}) from Eq.~(\ref{eq:KnFinal}) for $n=2,3,4,5$
and show the results in Tab.~\ref{tab:Universal-kernel}. In each
column, the kernel function denoted in the top row is obtained by
first multiplying the entries listed below it in the same column by
the common factor in the rightmost column and then taking the sum.
The symbols $\text{\ensuremath{\delta_{j}}}$ and $\text{\ensuremath{\Delta_{j}}}$
for $j=1,2,...,n-1$ which appear in Tab.~\ref{tab:Universal-kernel}
are defined by
\begin{eqnarray}
\text{\ensuremath{\delta_{j}}} & \ensuremath{\equiv} & \delta_{\Omega_{1}+\Omega_{2}+...+\Omega_{j},0},\label{eq:delta_j}\\
\text{\ensuremath{\Delta_{j}}} & \ensuremath{\equiv} & \Delta_{\Omega_{1}+\Omega_{2}+...+\Omega_{j}}\equiv\begin{cases}
0 & \textnormal{{if} \ensuremath{\Omega_{1}}+\ensuremath{\Omega_{2}}+...+\ensuremath{\Omega_{j}}=0}\\
\frac{1}{\Omega_{1}+\Omega_{2}+...+\Omega_{j}} & \textnormal{{if} \ensuremath{\Omega_{1}}+\ensuremath{\Omega_{2}}+...+\ensuremath{\Omega_{j}}\ensuremath{\ensuremath{\neq}0}}
\end{cases},\label{eq:Delta_j}
\end{eqnarray}
compare also to the previous section. As an example, for $n=2$ and
$n=3$ we obtain from Tab.~\ref{tab:Universal-kernel}
\begin{eqnarray}
K_{2}(\Omega_{1}) & = & -\Delta_{1}+\frac{\beta}{2}\delta_{1},\label{eq:K2}\\
K_{3}(\Omega_{1},\Omega_{2}) & = & +\Delta_{1}\Delta_{2}-\frac{\beta}{2}\delta_{1}\Delta_{2}-\Delta_{1}\delta_{2}\left(\frac{\beta}{2}+\Delta_{1}\right)+\delta_{1}\delta_{2}\frac{\beta}{2}\frac{\beta}{3},\label{eq:K3}
\end{eqnarray}
respectively. The rows of Tab.~\ref{tab:Universal-kernel} are organized
with respect to the number $a$ of factors $\delta_{l}$ in the summands.
Here, $a=0$ indicates the regular part and $a=1,2,...,n-1$ indicates
anomalous terms. There are $n-1$ \emph{choose $a$} anomalous terms
of order $a$. Our results are exact and go substantially beyond existing
expressions in the literature -- these are limited to $n\le3$ \citep{shvaika3pt2006}
or to fermionic $n=4$ \citep{hafermannSuperperturbationSolver2009,shvaika4pt2016,kuglerMultipointCorrelation2021}
with $a=0,1$ (and $a=2,3$ guaranteed to vanish, see below) or arbitrary
$n$ with $a=0$ \citep{kuglerMultipointCorrelation2021}. Alternative
expressions for the $n=3,4$ kernel functions with $a\leq1$ were
given in \citep{kuglerMultipointCorrelation2021}, but they are consistent
with our kernel functions as they yield the same correlation functions,
see the Appendix.
\noindent \begin{center}
\begin{table}
\noindent \begin{centering}
\begin{tabular}{l|l|l|l|l|l}
\#anom. & $K_{2}(\Omega_{1})$ & $K_{3}(\Omega_{1},\Omega_{2})$ & $K_{4}(\Omega_{1},\Omega_{2},\Omega_{3})$ & $K_{5}(\Omega_{1},\Omega_{2},\Omega_{3},\Omega_{4})$ & factor for entire row\tabularnewline
\hline 
$a=0$ & $-\Delta_{1}$ & $+\Delta_{1}\Delta_{2}$ & $-\Delta_{1}\Delta_{2}\Delta_{3}$ & $+\Delta_{1}\Delta_{2}\Delta_{3}\Delta_{4}$ & $1$\tabularnewline
\hline 
\multirow{4}{*}{$a=1$} & $+\delta_{1}$ & $-\delta_{1}\Delta_{2}$ & $+\delta_{1}\Delta_{2}\Delta_{3}$ & $-\delta_{1}\Delta_{2}\Delta_{3}\Delta_{4}$ & $\frac{\beta}{2}$\tabularnewline
 &  & $-\Delta_{1}\delta_{2}$ & $+\Delta_{1}\delta_{2}\Delta_{3}$ & $-\Delta_{1}\delta_{2}\Delta_{3}\Delta_{4}$ & $\frac{\beta}{2}+\Delta_{1}$\tabularnewline
 &  &  & $+\Delta_{1}\Delta_{2}\delta_{3}$ & $-\Delta_{1}\Delta_{2}\delta_{3}\Delta_{4}$ & $\frac{\beta}{2}+\Delta_{1}+\Delta_{2}$\tabularnewline
 &  &  &  & $-\Delta_{1}\Delta_{2}\Delta_{3}\delta_{4}$ & $\frac{\beta}{2}+\Delta_{1}+\Delta_{2}+\Delta_{3}$\tabularnewline
\hline 
\multirow{6}{*}{$a=2$} &  & $+\delta_{1}\delta_{2}$ & $-\delta_{1}\delta_{2}\Delta_{3}$ & $+\delta_{1}\delta_{2}\Delta_{3}\Delta_{4}$ & $\frac{\beta}{2}\frac{\beta}{3}$\tabularnewline
 &  & \multirow{1}{*}{} & $-\delta_{1}\Delta_{2}\delta_{3}$ & $+\delta_{1}\Delta_{2}\delta_{3}\Delta_{4}$ & $\frac{\beta}{2}\left(\frac{\beta}{3}+\Delta_{2}\right)$\tabularnewline
 &  &  & $-\Delta_{1}\delta_{2}\delta_{3}$ & $+\Delta_{1}\delta_{2}\delta_{3}\Delta_{4}$ & $\frac{\beta}{2}\frac{\beta}{3}+\Delta_{1}\left(\frac{\beta}{2}+\Delta_{1}\right)$\tabularnewline
 &  &  &  & $+\delta_{1}\Delta_{2}\Delta_{3}\delta_{4}$ & $\frac{\beta}{2}\left(\frac{\beta}{3}+\Delta_{2}+\Delta_{3}\right)$\tabularnewline
 &  &  &  & $+\Delta_{1}\delta_{2}\Delta_{3}\delta_{4}$ & $\frac{\beta}{2}\frac{\beta}{3}+\left(\Delta_{1}+\Delta_{3}\right)\left(\frac{\beta}{2}+\Delta_{1}\right)$\tabularnewline
 &  &  &  & $+\Delta_{1}\Delta_{2}\delta_{3}\delta_{4}$ & $\frac{\beta}{2}\frac{\beta}{3}+\left(\Delta_{1}+\Delta_{2}\right)\left(\frac{\beta}{2}+\Delta_{2}\right)+\Delta_{1}^{2}$\tabularnewline
\hline 
\multirow{4}{*}{$a=3$} &  &  & $+\delta_{1}\delta_{2}\delta_{3}$ & $-\delta_{1}\delta_{2}\delta_{3}\Delta_{4}$ & $\frac{\beta}{2}\frac{\beta}{3}\frac{\beta}{4}$\tabularnewline
 &  &  &  & $-\delta_{1}\delta_{2}\Delta_{3}\delta_{4}$ & $\frac{\beta}{2}\frac{\beta}{3}\left(\frac{\beta}{4}+\Delta_{3}\right)$\tabularnewline
 &  &  &  & $-\delta_{1}\Delta_{2}\delta_{3}\delta_{4}$ & $\frac{\beta}{2}\left(\frac{\beta}{3}\frac{\beta}{4}+\Delta_{2}\left(\frac{\beta}{3}+\Delta_{2}\right)\right)$\tabularnewline
 &  &  &  & $-\Delta_{1}\delta_{2}\delta_{3}\delta_{4}$ & $\frac{\beta}{2}\frac{\beta}{3}\frac{\beta}{4}+\Delta_{1}\left(\frac{\beta}{2}\frac{\beta}{3}+\Delta_{1}\left(\frac{\beta}{2}+\Delta_{1}\right)\right)$\tabularnewline
\hline 
$a=4$ &  &  &  & $+\delta_{1}\delta_{2}\delta_{3}\delta_{4}$ & $\frac{\beta}{2}\frac{\beta}{3}\frac{\beta}{4}\frac{\beta}{5}$\tabularnewline
\end{tabular}
\par\end{centering}
\caption{\label{tab:Universal-kernel}Universal kernel functions $K_{n}\bigl(\Omega_{1},...,\Omega_{n-1}\bigr)$
for $n=2,3,4,5$ defined in Eq.~(\ref{eq:KIncr}) and calculated
from Eq.~(\ref{eq:KnFinal}) in Sec.~\ref{sec:exact-expression}.
In each column, the kernel function in the top row is obtained by
first multiplying the entries listed below it in the same column by
the common factor in the rightmost column and then taking the sum,
see Eqns.~(\ref{eq:K2}) and (\ref{eq:K3}) as examples. The symbols
$\delta_{j}$ and $\Delta_{j}$ are defined in Eqns.~(\ref{eq:delta_j})
and (\ref{eq:Delta_j}). The rows are organized with respect to the
number $a$ of appearances of $\delta_{j}$, i.e.~the order of the
anomalous terms. }
\end{table}
\par\end{center}

In the case of purely fermionic correlators (all $A_{k}$ fermionic),
individual Matsubara frequencies $\omega_{k}$ cannot be zero. Thus
the complex numbers $\Omega_{k}^{\underline{a}\,\underline{b}}=i\omega_{k}+E_{\underline{a}}-E_{\underline{b}}$
of Eq.~(\ref{eq:Omega_k^ab}) always have a finite imaginary part,
regardless of the eigenenergies. In this case, only sums of an even
number of frequencies can be zero, and we can simplify $\delta_{1}=\delta_{3}=\delta_{5}=...=0$.
The expressions for the kernels in Tab.~\ref{tab:Universal-kernel},
now denoted by $K_{n}|_{F}$ for the fermionic case, simplify to
\begin{eqnarray}
K_{2}(\Omega_{1})|_{F} & = & -\Delta_{1},\label{eq:K2F}\\
K_{4}(\Omega_{1},\Omega_{2},\Omega_{3})|_{F} & = & \Delta_{1}\Delta_{3}\left[\delta_{2}\left(\frac{\beta}{2}+\Delta_{1}\right)-\Delta_{2}\right],\label{eq:K4F}\\
K_{6}(\Omega_{1},...,\Omega_{5})|_{F} & = & \Delta_{1}\Delta_{3}\Delta_{5}\biggl\{-\Delta_{2}\Delta_{4}-\delta_{2}\delta_{4}\left[\frac{\beta}{2}\frac{\beta}{3}+\left(\Delta_{1}+\Delta_{3}\right)\left(\frac{\beta}{2}+\Delta_{1}\right)\right]\label{eq:K6F}\\
 &  & +\delta_{4}\Delta_{2}\left(\frac{\beta}{2}+\Delta_{1}+\Delta_{2}+\Delta_{3}\right)+\delta_{2}\Delta_{4}\left(\frac{\beta}{2}+\Delta_{1}\right)\biggr\}.\nonumber 
\end{eqnarray}
This concludes the general part of this work. Next, we consider two
example systems frequently discussed in the condensed matter theory
literature. Using our formalism, we provide analytical forms of correlation
functions that to the best of our knowledge were not available before.

\section{Applications: Hubbard atom and Free Spin\label{sec:Applications}}

\subsection{Fermionic Hubbard atom}

The Hubbard atom (HA) describes an isolated impurity or otherwise
localized system with Hamiltonian
\begin{equation}
H=\epsilon(n_{\uparrow}+n_{\downarrow})+Un_{\uparrow}n_{\downarrow}-h(n_{\uparrow}-n_{\downarrow}),\label{eq:HAHamiltonian}
\end{equation}
see Fig.~\ref{fig:fig}(b) for a sketch. The HA corresponds to the
limit of vanishing system-bath coupling of the Anderson impurity model
(AIM), or vanishing hopping in the Hubbard model (HM). The particle
number operators $n_{\sigma}=d_{\sigma}^{\dagger}d_{\sigma}$ count
the number of fermionic particles with spin $\sigma\in\{\uparrow,\downarrow\}$,
each contributing an onsite energy $\epsilon$ shifted by an external
magnetic field $h$ in $z$-direction. An interaction energy $U$
is associated to double occupation.

Due to its simplicity and the four-dimensional Hilbert space, the
correlation functions for the HA can be found analytically using the
spectral representation. It is therefore often used for benchmarking
\citep{krienParquetlikeEquations2019,krienTilingTriangles2021,kapplNonlinearResponses2022}.
The presence of the interaction term leads to a non-vanishing $n=4$
one-line irreducible vertex function. The HA serves as an important
reference point to study and interpret properties of the AIM and HM
beyond the one-particle level, for example divergences of two-line
irreducible vertex functions \citep{schaferNonperturbativeLandscape2016,thunstromAnalyticalInvestigation2018,chalupaDivergencesIrreducible2018,pelzHighlyNonperturbative2023}
and signatures of the local moment formation in generalized susceptibilities
\citep{chalupaFingerprintsLocal2021,adlerNonperturbativeIntertwining2022}.
Using the fermionic kernels in Eqns.~(\ref{eq:K2F}) and (\ref{eq:K4F}),
we have checked that our formalism reproduces the results for the
2-point and 4-point correlators given in Refs.~\citep{hafermannSuperperturbationSolver2009,rohringerNewRoutes2013,kuglerMultipointCorrelation2021}
for half-filling, $\epsilon=-U/2$ and $h=0$. 

Correlation functions including bosonic operators describe the asymptotic
behaviour of the $n=4$ fermion vertex for large frequencies \citep{wentzellHighfrequencyAsymptotics2020}
or the interaction of electrons by the exchange of an effective boson
\citep{krienSinglebosonExchange2019,gieversMultiloopFlow2022}. These
relations involve correlation functions of two bosonic operators or
of one bosonic and two fermionic operators, giving rise to expressions
possibly anomalous in at most one frequency argument, i.e. $a\leq1$.

For the HA, AIM and HM, bosonic correlation functions for $n>2$ have
not been considered thoroughly so far. Only recently, steps in this
direction were taken, particularly in the context of non-linear response
theory \citep{kapplNonlinearResponses2022}. The response of a system
to first and second order in an external perturbation is described
by $2$- and $3$-point correlation functions, respectively. For the
HA, physically motivated perturbations affect the onsite energy via
a term $\delta_{\epsilon}n$ or take the form of a magnetic field
$\boldsymbol{\delta}_{h}\cdot\mathbf{S}$. Here, the parameters $\delta_{\varepsilon}$
and $\boldsymbol{\delta}_{h}$ denote the strength of the perturbation
and we define
\begin{equation}
\begin{aligned}n & =n_{\uparrow}+n_{\downarrow}\end{aligned}
,\quad S^{x}=\frac{1}{2}\left(d_{\uparrow}^{\dagger}d_{\downarrow}+d_{\downarrow}^{\dagger}d_{\uparrow}\right),\quad S^{y}=\frac{-i}{2}\left(d_{\uparrow}^{\dagger}d_{\downarrow}-d_{\downarrow}^{\dagger}d_{\uparrow}\right),\quad S^{z}=\frac{1}{2}\left(n_{\uparrow}-n_{\downarrow}\right).\label{eq:HA_nSxSySz}
\end{equation}
The resulting changes of the expectation values of the density or
magnetization in arbitrary direction are described in second order
of the perturbation by the connected parts of the correlation functions
$G_{A_{1}A_{2}A_{3}}(\tau_{1},\tau_{2},\tau_{3})$, with $A_{i}\in\{n,S_{x},S_{y},S_{z}\}$,
where the time-ordered expectation value is evaluated with respect
to the unperturbed system (\ref{eq:HAHamiltonian}) and Fourier transformed
to the frequencies of interest. These objects have been studied numerically
in Ref.~ \citep{kapplNonlinearResponses2022}. In the following,
we give explicit, analytic expressions of the full correlation functions
$G_{A_{1}A_{2}A_{3}}(\omega_{1},\omega_{2})$ (i.e.~including disconnected
parts), for arbitrary parameters $\epsilon$, $U$ and $h$ and for
all possible operator combinations using the (bosonic) kernel function
$K_{3}$, see Eq.~(\ref{eq:K3}). To the best of our knowledge, these
expressions have not been reported before.

The eigenstates of the HA Hamiltonian (\ref{eq:HAHamiltonian}) {[}see
Fig.~\ref{fig:fig}(b){]} describe an empty ($\vert0\rangle$), singly
occupied ($d_{\uparrow}^{\dagger}\vert0\rangle=\vert\uparrow\rangle$,
$d_{\downarrow}^{\dagger}\vert0\rangle=\vert\downarrow\rangle$) or
doubly occupied ($d_{\uparrow}^{\dagger}d_{\downarrow}^{\dagger}\vert0\rangle=\vert\uparrow\downarrow\rangle$)
impurity with eigenenergies $E_{0}=0$, $E_{\uparrow}=\epsilon-h$,
$E_{\downarrow}=\epsilon+h$ and $E_{\uparrow\downarrow}=2\epsilon+U$,
respectively. The partition function is $Z=1+e^{-\beta(\epsilon-h)}+e^{-\beta(\epsilon+h)}+e^{-\beta(2\epsilon+U)}.$
We define
\begin{equation}
s=\frac{e^{-\beta\epsilon}}{Z}\sinh(\beta h),\quad c=\frac{e^{-\beta\epsilon}}{Z}\cosh(\beta h),
\end{equation}
and obtain all non-vanishing bosonic 3-point correlation functions
(where $\omega_{3}=-\omega_{1}-\omega_{2}$):
\begin{align}
G_{nnn}(\omega_{1},\omega_{2}) & =2\beta^{2}\delta_{\omega_{1}}\delta_{\omega_{2}}\left(\frac{4e^{-\beta(2\epsilon+U)}}{Z}+c\right),\\
G_{nnS^{z}}(\omega_{1},\omega_{2}) & =\beta^{2}\delta_{\omega_{1}}\delta_{\omega_{2}}s,\\
G_{nS^{x}S^{y}}(\omega_{1},\omega_{2}) & =-\beta\delta_{\omega_{1}}s\frac{\omega_{2}}{\omega_{2}^{2}+4h^{2}},\\
G_{nS^{x}S^{x}}(\omega_{1},\omega_{2})=G_{nS^{y}S^{y}}(\omega_{1},\omega_{2}) & =2\beta\delta_{\omega_{1}}\frac{h\ s}{\omega_{2}^{2}+4h^{2}},\\
G_{nS^{z}S^{z}}(\omega_{1},\omega_{2}) & =\frac{\beta^{2}}{2}\delta_{\omega_{1}}\delta_{\omega_{2}}c,\\
G_{S^{z}S^{x}S^{x}}(\omega_{1},\omega_{2})=G_{S^{z}S^{y}S^{y}}(\omega_{1},\omega_{2}) & =-s\frac{\omega_{2}\omega_{3}+4h^{2}}{(\omega_{2}^{2}+4h^{2})(\omega_{3}^{2}+4h^{2})}+\beta\delta_{\omega_{1}}\frac{h\ c}{\omega_{2}^{2}+4h^{2}},\\
G_{S^{z}S^{z}S^{z}}(\omega_{1},\omega_{2}) & =\frac{\beta^{2}}{4}\delta_{\omega_{1}}\delta_{\omega_{2}}s,\\
G_{S^{x}S^{y}S^{z}}(\omega_{1},\omega_{2}) & =2h\ s\frac{\omega_{1}-\omega_{2}}{(\omega_{1}^{2}+4h^{2})(\omega_{2}^{2}+4h^{2})}-\frac{\beta}{2}\delta_{\omega_{3}}c\frac{\omega_{1}}{\omega_{1}^{2}+4h^{2}}.\label{eq:GSxSySz}
\end{align}
We observe that each conserved quantity, in this case $n$ and $S_{z}$,
contributes an anomalous term $\propto\delta_{\omega_{k}}$in its
respective frequency argument $\omega_{k}$. If an operator $A_{k}$
is conserved $[H,A_{k}]=0$, the basis over which we sum in Eq.~(\ref{eq:Spectral_Gw})
can be chosen such that both $H$ and $A_{k}$ are diagonal, $A_{k}^{\underline{1}\,\underline{2}}=A_{k}^{\underline{1}\,\underline{1}}\delta_{\underline{1},\underline{2}}$.
If $A_{k}^{\underline{1}\,\underline{1}}\neq0$ for some state $\underline{1}$
the vanishing eigenenergy difference leads to the appearance of an
anomalous contribution. If the operators in the correlator additionally
commute with each other, in our case for example $[n,S^{z}]=0$, there
exists a basis in which all operators and the Hamiltonian are diagonal,
giving rise to correlation functions anomalous in all frequency arguments.

In the limit of vanishing field $h\rightarrow0$, we introduce an
additional degeneracy $E_{\uparrow}=E_{\downarrow}=\epsilon$ in the
system, potentially resulting in additional anomalous contributions.
The corresponding correlation functions can then be obtained in two
ways. Either we recompute them using the kernel function $K_{3}$
or we take appropriate limits, for example
\begin{equation}
\lim_{h\rightarrow0}\frac{h\ \sinh(\beta h)}{\omega_{k}^{2}+4h^{2}}=\frac{\beta}{4}\delta_{\omega_{k}},
\end{equation}
resulting in
\begin{align}
G_{nnn}(\omega_{1},\omega_{2}) & =\beta^{2}\delta_{\omega_{1}}\delta_{\omega_{2}}\frac{2(4e^{-\beta(2\epsilon+U)}+e^{-\beta\epsilon})}{Z},\\
G_{nS^{\alpha}S^{\alpha}}(\omega_{1},\omega_{2}) & =\beta^{2}\delta_{\omega_{1}}\delta_{\omega_{2}}\frac{e^{-\beta\epsilon}}{2Z}\quad(\alpha\in\{x,y,z\}),\\
G_{S^{x}S^{y}S^{z}}(\omega_{1},\omega_{2}) & =\beta\frac{e^{-\beta\epsilon}}{2Z}(-\delta_{\omega_{1}}\Delta_{\omega_{2}}+\delta_{\omega_{2}}\Delta_{\omega_{1}}-\delta_{\omega_{1}+\omega_{2}}\Delta_{\omega_{1}}),\label{eq:GSxSySz(h=00003D0)}
\end{align}
with all other correlation functions vanishing. As already pointed
out in Ref.~\citep{kapplNonlinearResponses2022}, only the last correlation
function retains a nontrivial frequency dependence due to non-commuting
operators. 

\subsection{Free spin $S$}

We now consider correlation functions of a free spin of length $S$,
without a magnetic field, so that temperature $T=1/\beta$ is the
only energy scale. The operators $\{S^{\alpha}\}_{\alpha=x,y,z}$
fulfill $S^{x}S^{x}+S^{y}S^{y}+S^{z}S^{z}=S(S+1)$ and the SU(2) algebra
$\left[S^{\alpha_{1}},S^{\alpha_{2}}\right]=i\sum_{\alpha_{3}=\{x,y,z\}}\epsilon^{\alpha_{1}\alpha_{2}\alpha_{3}}S^{\alpha_{3}}$,
thus they are bosonic. Since the Hamiltonian vanishes and therefore
all eigenenergies are zero, every $\Omega_{k}^{\underline{a}\,\underline{b}}$
in the spectral representation (\ref{eq:Spectral_Gw}) can vanish
and a proper treatment of all anomalous terms is essential. As the
Heisenberg time dependence is trivial, $S^{\alpha}(\tau)=S^{\alpha}$,
the non-trivial frequency dependence of the correlators, which can
be can be non-vanishing at any order $n$, derives solely from the
action of imaginary time-ordering.

The correlators are required, for example, as the non-trivial initial
condition for the spin-fRG recently suggested by Kopietz et al., Refs.~\citep{kriegExactRenormalization2019,gollSpinFunctional2019,gollZeromagnonSound2020,tarasevychDissipativeSpin2021,tarasevychSpinFunctional2022}.
However, for $n>3$ they are so far only partially available: They
are either given for restricted frequency combinations, or for the
purely classical case $S^{\alpha_{1}}=S^{\alpha_{2}}=...=S^{\alpha_{n}}$
where the SU(2) algebra does not matter, or for finite magnetic field
via an equation of motion \citep{gollSpinFunctional2019} or diagrammatic
approach \citep{vaksThermodynamicsIdeal,vaksSpinWaves1968}.

We define the spin raising and lowering operators,
\begin{equation}
S^{\pm}=\left(S^{x}\pm iS^{y}\right)/\sqrt{2},\label{eq:Spm}
\end{equation}
which have to appear in pairs for a non-vanishing correlator due to
spin-rotation symmetry. As for the HA, we do not consider connected
correlators in this work for brevity. The classical $S^{z}$-correlator
can be found from its generating functional with source field $h$
\citep{kriegExactRenormalization2019},
\begin{eqnarray}
\mathcal{G}\left(y=\beta h\right) & = & \frac{\sinh\left[y(S+1/2)\right]}{(2S+1)\sinh\left[y/2\right]},\\
\left\langle (S^{z})^{l}\right\rangle  & = & \underset{y\rightarrow0}{\mathrm{lim}}\partial_{y}^{l}\mathcal{G}(y)\equiv b_{l-1},\label{eq:bl-1}
\end{eqnarray}
for example $b_{1}=\frac{S}{3}(S+1)$ and $b_{3}=\frac{S}{15}\left(3S^{3}+6S^{2}+2S-1\right)$
and vanishing $b_{l}$ for even $l$. For all other correlators involving
$\alpha_{k}=\pm$, we adapt Eq.~(\ref{eq:Spectral_Gw}) for the free
spin case,
\begin{equation}
G_{S^{\alpha_{1}}S^{\alpha_{2}}...S^{\alpha_{n}}}\left(\omega_{1},...,\omega_{n-1}\right)=\sum_{p\in S_{n}}\left\langle S^{\alpha_{p(1)}}S^{\alpha_{p(2)}}...S^{\alpha_{p(n)}}\right\rangle K_{n}\left(i\omega_{p(1)},i\omega_{p(2)},...,i\omega_{p(n-1)}\right),\label{eq:Gw_spectral_freeSpin}
\end{equation}
where we made use of the fact that all eigenenergies are zero and
the Heisenberg time evolution is trivial. It is convenient to evaluate
the equal-time correlators in Eq.~(\ref{eq:Gw_spectral_freeSpin})
as
\begin{equation}
\left\langle S^{\alpha_{1}}S^{\alpha_{2}}...S^{\alpha_{n}}\right\rangle =\frac{1}{2S+1}\sum_{m=-S}^{S}\left\langle m\right|S^{\alpha_{1}}S^{\alpha_{2}}...S^{\alpha_{n}}\left|m\right\rangle \equiv\frac{1}{2S+1}\sum_{m=-S}^{S}\sum_{l=0}^{n}p_{l}m^{l}=p_{0}+\sum_{l=2}^{n}p_{l}b_{l-1}
\end{equation}
where in the last step we used Eq.~(\ref{eq:bl-1}). We find the
real expansion coefficients $\{p_{l}\}_{l=0,1,...,n}$ iteratively
by moving through the string $\alpha_{1}\alpha_{2}...\alpha_{n}$
from the right and start from $p_{l}=\delta_{0,l}$. Based on the
$S^{z}$ eigenstates $\{\bigl|m\bigr\rangle\}_{m=-S,...,S-1,S}$ we
obtain the iteration rules from $S^{z}\bigl|m\bigr\rangle=m\bigl|m\bigr\rangle$
and $S^{\pm}\bigl|m\bigr\rangle=\sqrt{1/2}\sqrt{S(S+1)-m(m\pm1)}\bigl|m\pm1\bigr\rangle$.
We define an auxiliary integer $c$ that keeps track of the intermediate
state $\left|m+c\right\rangle $, initially $c=0$. Depending on the
$\alpha_{j}$ that we find in step $j=n,n-1...,1$ we take one of
the following actions: (i) For $\alpha_{j}=z$, we update $p_{l}\leftarrow p_{l-1}+cp_{l}\;\forall l$
and leave $c$ unchanged. It is understood that $p_{l<0}=0$. (ii)
For $\alpha_{j}=+$, we combine the square-root factor brought by
the raising operator with the factor that comes from the necessary
$\alpha_{j^{\prime}}=-$ at another place in the string. We replace
$p_{l}\leftarrow-\frac{1}{2}p_{l-2}-\frac{2c+1}{2}p_{l-1}+\left(\frac{3}{2}b_{1}-c\frac{c+1}{2}\right)p_{l}\;\forall l$
and then let $c\leftarrow c+1$. (iii) For $\alpha_{j}=-$, we update
$c\leftarrow c-1$ and keep $p_{l}$ unchanged, $p_{l}\leftarrow p_{l}\;\forall l$.
\begin{table}
\noindent \begin{centering}
\begin{tabular}{c|l}
$n=2$ & $G_{S^{+}S^{-}}(\omega)=G_{S^{z}S^{z}}(\omega)=\beta\delta_{\omega}b_{1}$\tabularnewline
\hline 
$n=3$ & $G_{S^{+}S^{-}S^{z}}(\omega_{1},\omega_{2})=\beta b_{1}(-\delta_{\omega_{1}}\Delta_{i\omega_{2}}+\delta_{\omega_{2}}\Delta_{i\omega_{1}}+\delta_{\omega_{1}+\omega_{2}}\Delta_{i\omega_{2}})=-iG_{S^{x}S^{y}S^{z}}(\omega_{1},\omega_{2})$\tabularnewline
\hline 
\multirow{4}{*}{$n=4$} & $G_{S^{z}S^{z}S^{z}S^{z}}\left(\omega_{1},\omega_{2},\omega_{3}\right)=\delta_{\omega_{1}}\delta_{\omega_{2}}\delta_{\omega_{3}}\beta^{3}b_{3}$\tabularnewline
 & $G_{S^{+}S^{+}S^{-}S^{-}}\left(\omega_{1},\omega_{2},\omega_{3}\right)=\beta b_{1}[2\times\delta_{\omega_{1}}\delta_{\omega_{2}}\delta_{\omega_{3}}\times\frac{\beta^{2}}{5}\left(3b_{1}-\frac{1}{3}\right)+r]$\tabularnewline
 & $G_{S^{+}S^{-}S^{z}S^{z}}\,\left(\omega_{1},\omega_{2},\omega_{3}\right)\,=\beta b_{1}[1\times\delta_{\omega_{1}}\delta_{\omega_{2}}\delta_{\omega_{3}}\times\frac{\beta^{2}}{5}\left(3b_{1}-\frac{1}{3}\right)-r]$\tabularnewline
 & $r=\Delta_{i\omega_{1}}\Delta_{i\text{\ensuremath{\omega_{2}}}}\left(\delta_{\omega_{1}+\omega_{3}}+\delta_{\omega_{2}+\omega_{3}}-\delta_{\omega_{3}}-\delta_{\omega_{4}}\right)-\left(\delta_{\omega_{1}}\Delta_{i\text{\ensuremath{\omega_{2}}}}^{2}+\delta_{\omega_{2}}\Delta_{i\text{\ensuremath{\omega_{1}}}}^{2}\right)\left(\delta_{\omega_{3}}+\delta_{\text{\ensuremath{\omega_{4}}}}\right)-\Delta_{i\omega_{3}}\Delta_{i\omega_{4}}\left(\delta_{\omega_{1}}+\delta_{\omega_{2}}\right)$\tabularnewline
\end{tabular}
\par\end{centering}
\caption{\label{tab:spin-G}Matsubara correlation functions for a free spin-$S$
up to order $n=4$. Here, $\omega_{4}=-\omega_{1}-\omega_{2}-\omega_{3}$.}
\end{table}

Our final results for the free spin correlators are reported in Tab.~\ref{tab:spin-G}.
We reproduce the known spin correlators for $n=2,3$ and determine
the non-classical correlators $G_{S^{+}S^{+}S^{-}S^{-}}$ and $G_{S^{+}S^{-}S^{z}S^{z}}$
at order $n=4$, which to the best of our knowledge were not available
in the literature \citep{thanksAR}. We also confirmed the classical
result for $G_{S^{z}S^{z}S^{z}S^{z}}$, which in our full quantum
formalism requires some non-trivial cancellations. To arrive at our
results, we used the identity
\begin{eqnarray}
\Delta_{a+b}\left(\Delta_{a}+\Delta_{b}\right)-\Delta_{a}\Delta_{b} & = & \delta_{a}\Delta_{b}^{2}+\delta_{b}\Delta_{a}^{2}-\delta_{a+b}\Delta_{a}\Delta_{b}.\label{eq:DeltaDelta}
\end{eqnarray}
We finally comment on the relation between the $n=3$ free spin-$S$correlator
$G_{S^{+}S^{-}S^{z}}$ from Tab.~(\ref{tab:spin-G}) and the result
for $G_{S^{x}S^{y}S^{z}}$ found for the zero-field limit of the HA
in Eq.~(\ref{eq:GSxSySz(h=00003D0)}). The operators $S^{x,y,z}$
for the Hubbard model {[}c.f.~Eq.~(\ref{eq:HA_nSxSySz}){]} project
to the singly-occupied $S=1/2$ subspace spanned by the states $\bigl|\uparrow\bigr\rangle,\bigl|\downarrow\bigr\rangle$.
Thus, using $G_{S^{x}S^{y}S^{z}}=iG_{S^{+}S^{-}S^{z}}$ and specializing
the free spin result from Tab.~(\ref{tab:spin-G}) to $S=1/2$ (where
$b_{1}=1/4$) we find agreement with the HA result (\ref{eq:GSxSySz(h=00003D0)})
up to the factor $2e^{-\beta\epsilon}/Z$. This factor represents
the expectation value of the projector to the singly-occupied sector
in the HA Hilbert space and goes to unity in the local-moment regime.

\section{Conclusion\label{sec:Conclusion}}

In summary, we have provided exact universal kernel functions for
the spectral representation of the $n$-point Matsubara correlator.
Our results are an efficient alternative to equation-of-motion approaches
which often have difficulties to capture anomalous terms related to
conserved or commuting operators. We expect our results to be useful
for various benchmarking applications, as starting points for emerging
many-body methods and for unraveling the physical interpretation of
$n$-point functions in various settings. Our results also apply in
the limit $T\rightarrow0$ where the formally divergent anomalous
contributions are to be understood as $\beta\delta_{\omega,0}\rightarrow2\pi\delta(\omega)$.
Some of these Dirac delta-functions will vanish after subtracting
the disconnected contributions, others indicate truely divergent susceptibilities
like the $1/T$ Curie law for the spin-susceptiblity of the Hubbard
atom in the local moment regime \citep{rohringerNewRoutes2013}. Although
our work has focused on imaginary frequency (Matsubara) correlators,
with analytical expressions now at hand, it is also interesting to
study the intricacies of analytical continuation to real frequencies
and thus to further explore the connection of Matsubara and Keldysh
correlators \citep{geAnalyticalContinuation}.
\begin{acknowledgments}
We acknowledge useful discussions with Karsten Held, Friedrich Krien,
Seung-Sup Lee, Peter Kopietz, Fabian Kugler, Nepomuk Ritz, Georg Rohringer,
Andreas Rückriegel. We thank Andreas Rückriegel for sharing unpublished
results on 4-point free spin correlators and pointing out further
simplifications of the analytical expressions. BS and BS are supported
by a MCQST-START fellowship. We acknowledge funding from the International
Max Planck Research School for Quantum Science and Technology (IMPRSQST)
for JH, from the Deutsche Forschungsgemeinschaft under Germany\textquoteright s
Excellence Strategy EXC-2111 (Project No. 390814868), and from the
Munich Quantum Valley, supported by the Bavarian state government
with funds from the Hightech Agenda Bayern Plus.
\end{acknowledgments}

\section*{Appendix: Equivalence to convention of Ref.~{[}21{]}\label{app:Equivalence}}

In Ref.~\citep{kuglerMultipointCorrelation2021} by Kugler, Lee and
von Delft (KLD), only regular ($a=0$) and anomalous terms of order
$a=1$ have been considered for $n=3,4$. The corresponding kernel
functions were derived from only $(n-1)!$ permutations by setting
$\tau_{n}=0$ and $\tau_{i\neq n}>0$, but still applied to all $n!$
permutations to obtain the correlation functions. For $n=3$, the
resulting kernel function (Eq.~(46) in Ref.~\citep{kuglerMultipointCorrelation2021})
reads
\begin{equation}
K_{3,\textnormal{KLD}}(\Omega_{1},\Omega_{2})=\Delta_{1}\Delta_{2}-\Delta_{1}\delta_{2}\frac{1}{2}\left(\beta+\Delta_{1}\right)-\delta_{1}\Delta_{2}\frac{1}{2}\left(\beta+\Delta_{2}\right).\label{eq:KLD3pt-1}
\end{equation}
This can be compared to the corresponding kernel function for $n=3$
found in our Eq.~(\ref{eq:K3}) truncated to $a\le1$,
\begin{equation}
K_{3}^{a\le1}(\Omega_{1},\Omega_{2})=\Delta_{1}\Delta_{2}-\Delta_{1}\delta_{2}\left(\frac{\beta}{2}+\Delta_{1}\right)-\frac{\beta}{2}\delta_{1}\Delta_{2}.\label{eq:K_3^a<=00003D1}
\end{equation}

Both approaches are equally valid and should yield the same correlation
functions (consistently discarding terms with $a=2$), yet the kernel
functions are obviously different. To resolve this issue, we define
the difference of the kernel functions
\begin{equation}
K_{3,\textnormal{diff}}(\Omega_{1},\Omega_{2})=K_{3,\textnormal{KLD}}(\Omega_{1},\Omega_{2})-K_{3}^{a\le1}(\Omega_{1},\Omega_{2})=\frac{1}{2}\left(\Delta_{1}^{2}\delta_{2}-\delta_{1}\Delta_{2}^{2}\right)
\end{equation}
and show that the corresponding contributions to the correlation function
vanishes when summed over cyclically related permutations $p=123,231,312$.
These contributions are given by
\begin{equation}
\begin{aligned} & \frac{1}{Z}\sum_{p=123,231,312}\zeta(p)\sum_{\underline{1}\underline{2}\underline{3}}e^{-\beta E_{\underline{1}}}A_{p(1)}^{\underline{1}\underline{2}}A_{p(2)}^{\underline{2}\underline{3}}A_{p(3)}^{\underline{3}\underline{1}}K_{3,\textnormal{diff}}(\Omega_{p(1)}^{\underline{1}\,\underline{2}},\Omega_{p(2)}^{\underline{2}\,\underline{3}})\\
 & =\frac{\zeta(123)}{2Z}\sum_{\underline{1}\underline{2}\underline{3}}e^{-\beta E_{\underline{1}}}A_{1}^{\underline{1}\underline{2}}A_{2}^{\underline{2}\underline{3}}A_{3}^{\underline{3}\underline{1}}\left(\frac{(1-\delta_{\omega_{1}}\delta_{E_{\underline{1}}-E_{\underline{2}}})\delta_{\omega_{1}+\omega_{2}}\delta_{E_{\underline{1}}-E_{\underline{3}}}}{(i\omega_{1}+E_{\underline{1}}-E_{\underline{2}})^{2}}-\frac{\delta_{\omega_{1}}\delta_{E_{\underline{1}}-E_{\underline{2}}}(1-\delta_{\omega_{1}+\omega_{2}}\delta_{E_{\underline{1}}-E_{\underline{3}}})}{(i\omega_{1}+i\omega_{2}+E_{\underline{1}}-E_{\underline{3}})^{2}}\right)\\
 & +\frac{\zeta(231)}{2Z}\sum_{\underline{1}\underline{2}\underline{3}}e^{-\beta E_{\underline{1}}}A_{2}^{\underline{1}\underline{2}}A_{3}^{\underline{2}\underline{3}}A_{1}^{\underline{3}\underline{1}}\left(\frac{(1-\delta_{\omega_{2}}\delta_{E_{\underline{1}}-E_{\underline{2}}})\delta_{\omega_{2}+\omega_{3}}\delta_{E_{\underline{1}}-E_{\underline{3}}}}{(i\omega_{2}+E_{\underline{1}}-E_{\underline{2}})^{2}}-\frac{\delta_{\omega_{2}}\delta_{E_{\underline{1}}-E_{\underline{2}}}(1-\delta_{\omega_{2}+\omega_{3}}\delta_{E_{\underline{1}}-E_{\underline{3}}})}{(i\omega_{2}+i\omega_{3}+E_{\underline{1}}-E_{\underline{3}})^{2}}\right)\\
 & +\frac{\zeta(312)}{2Z}\sum_{\underline{1}\underline{2}\underline{3}}e^{-\beta E_{\underline{1}}}A_{3}^{\underline{1}\underline{2}}A_{1}^{\underline{2}\underline{3}}A_{2}^{\underline{3}\underline{1}}\left(\frac{(1-\delta_{\omega_{3}}\delta_{E_{\underline{1}}-E_{\underline{2}}})\delta_{\omega_{3}+\omega_{1}}\delta_{E_{\underline{1}}-E_{\underline{3}}}}{(i\omega_{3}+E_{\underline{1}}-E_{\underline{2}})^{2}}-\frac{\delta_{\omega_{3}}\delta_{E_{\underline{1}}-E_{\underline{2}}}(1-\delta_{\omega_{3}+\omega_{1}}\delta_{E_{\underline{1}}-E_{\underline{3}}})}{(i\omega_{3}+i\omega_{1}+E_{\underline{1}}-E_{\underline{3}})^{2}}\right).
\end{aligned}
\label{eq:Diffsum}
\end{equation}
Considering the second term of permutation $p=312$ and renaming the
summation variables $\underline{2}\rightarrow\underline{1}$, $\underline{3}\rightarrow\underline{2}$,
$\underline{1}\rightarrow\underline{3}$ yields
\begin{equation}
\begin{aligned} & -\frac{\zeta(312)}{2Z}\sum_{\underline{1}\underline{2}\underline{3}}e^{-\beta E_{\underline{1}}}A_{3}^{\underline{1}\underline{2}}A_{1}^{\underline{2}\underline{3}}A_{2}^{\underline{3}\underline{1}}\frac{\delta_{\omega_{3}}\delta_{E_{\underline{1}}-E_{\underline{2}}}(1-\delta_{\omega_{3}+\omega_{1}}\delta_{E_{\underline{1}}-E_{\underline{3}}})}{i(\omega_{3}+\omega_{1})+E_{\underline{1}}-E_{\underline{3}}}\\
 & =-\frac{\zeta(312)}{2Z}\sum_{\underline{1}\underline{2}\underline{3}}e^{-\beta E_{\underline{3}}}A_{1}^{\underline{1}\underline{2}}A_{2}^{\underline{2}\underline{3}}A_{3}^{\underline{3}\underline{1}}\frac{\delta_{\omega_{3}}\delta_{E_{\underline{3}}-E_{\underline{1}}}(1-\delta_{\omega_{3}+\omega_{1}}\delta_{E_{\underline{3}}-E_{\underline{2}}})}{(i\omega_{3}+i\omega_{1}+E_{\underline{3}}-E_{\underline{2}})^{2}}\\
 & =-\frac{\zeta(123)}{2Z}\sum_{\underline{1}\underline{2}\underline{3}}e^{-\beta E_{\underline{1}}}A_{1}^{\underline{1}\underline{2}}A_{2}^{\underline{2}\underline{3}}A_{3}^{\underline{3}\underline{1}}\frac{\delta_{\omega_{1}+\omega_{2}}\delta_{E_{\underline{1}}-E_{\underline{3}}}(1-\delta_{\omega_{1}}\delta_{E_{\underline{1}}-E_{\underline{2}}})}{(i\omega_{1}+E_{\underline{1}}-E_{\underline{2}})^{2}},
\end{aligned}
\end{equation}
where we used $\omega_{3}=-\omega_{1}-\omega_{2}$ and the fact that
$\delta_{\omega_{3}}$ enforces the third operator to be bosonic,
such that $\zeta(312)=\zeta(123)$. This term exactly cancels the
first contribution of permutation $p=123$ in (\ref{eq:Diffsum}).
Repeating similar steps for the remaining terms, we find that the
the second term of $p=123$ and the first term of $p=231$ as well
as the second term of $p=231$ and the first term of $p=312$ cancel,
leading to 
\begin{equation}
\frac{1}{Z}\sum_{p\in\{123,231,312\}}\zeta(p)\sum_{\underline{1}\underline{2}\underline{3}}e^{-\beta E_{\underline{1}}}A_{p(1)}^{\underline{1}\underline{2}}A_{p(2)}^{\underline{2}\underline{3}}A_{p(3)}^{\underline{3}\underline{1}}K_{3,\textnormal{diff}}(\Omega_{p(1)}^{\underline{1}\underline{2}},\Omega_{p(2)}^{\underline{2}\underline{3}})=0.
\end{equation}
Similarly, summing over the second set of cyclically related permutations
$p=132,213,321$ leads to a vanishing result, leading to the conclusion
that
\begin{equation}
\frac{1}{Z}\sum_{p\in S_{3}}\zeta(p)\sum_{\underline{1}\underline{2}\underline{3}}e^{-\beta E_{\underline{1}}}A_{p(1)}^{\underline{1}\underline{2}}A_{p(2)}^{\underline{2}\underline{3}}A_{p(3)}^{\underline{3}\underline{1}}K_{3,\textnormal{diff}}(\Omega_{p(1)}^{\underline{1}\underline{2}},\Omega_{p(2)}^{\underline{2}\underline{3}})=0.
\end{equation}
Thus we have shown that both kernel functions in Eqns.~(\ref{eq:KLD3pt-1})
and (\ref{eq:K_3^a<=00003D1}) are equivalent as they yield the same
correlation functions after summing over all permutations. The same
statement holds true for case of $n=4$ and $a=1$.

\bibliographystyle{unsrtnat}
\bibliography{library}

\end{document}